# Data Fusion on Motion and Magnetic Sensors embedded on Mobile Devices for the Identification of Activities of Daily Living

Ivan Miguel Pires[1,2,3], Nuno M. Garcia[1,3,4], Nuno Pombo[1,3,4], Francisco Flórez-Revuelta[5] and Susanna Spinsante[6]

[1]Instituto de Telecomunicações, Universidade da Beira Interior, Covilhã, Portugal
[2]Altranportugal, Lisbon, Portugal
[3]ALLab - Assisted Living Computing and Telecommunications Laboratory, Computer Science Department, Universidade da Beira Interior, Covilhã, Portugal
[4]Universidade Lusófona de Humanidades e Tecnologias, Lisbon, Portugal
[5]Department of Computer Technology, Universidad de Alicante, Spain
[6]Università Politecnica delle Marche, Ancona, Italy

impires@it.ubi.pt, ngarcia@di.ubi.pt, ngpombo@di.ubi.pt, francisco.florez@ua.es, s.spinsante@univpm.it

**Abstract**

Several types of sensors have been available in off-the-shelf mobile devices, including motion, magnetic, vision, acoustic, and location sensors. This paper focuses on the fusion of the data acquired from motion and magnetic sensors, *i.e.,* accelerometer, gyroscope and magnetometer sensors, for the recognition of Activities of Daily Living (ADL) using pattern recognition techniques. The system developed in this study includes data acquisition, data processing, data fusion, and artificial intelligence methods. Artificial Neural Networks (ANN) are included in artificial intelligence methods, which are used in this study for the recognition of ADL. The purpose of this study is the creation of a new method using ANN for the identification of ADL, comparing three types of ANN, in order to achieve results with a reliable accuracy. The best accuracy was obtained with Deep Learning, which, after the application of the $L_2$ regularization and normalization techniques on the sensors' data, reports an accuracy of 89.51%.

## 1. Introduction

Off-the-shelf mobile devices have several sensors available, which are capable for the acquisition of several physical and physiological parameters [1], including the accelerometer, magnetometer, and gyroscope sensors, allowing the recognition of Activities of Daily Living (ADL) [2]. The correct identification of ADL is one of the stages for the development of a personal digital life coach [3], which can be used in several areas, including sport, geriatrics, among others.

For the use of several sensors in the development of a method for the recognition of ADL, data fusion techniques should be used before the application of the artificial intelligence techniques. This paper focuses on the use of motion and magnetic sensors available on mobile devices, where the most commonly available are the accelerometer, the magnetometer, and the gyroscope, proposing the recognition of ADL with movement, including running, walking, walking on stairs, and standing. The architecture for the method for the recognition of ADL was proposed in [4-6], which is composed by data acquisition, data processing, data fusion, and artificial intelligence methods. Taking in account that the data acquired from the sensors is fulfilled, the data processing methods are forked in two types of methods, such as data cleaning, and features extraction methods. After the feature extraction, the data fusion and artificial intelligence methods are commonly applied at the same time.

Recently, several studies have been performed for the recognition of ADL using several sensors [7-12], proving the reliability of the use of Artificial Neural Networks (ANN) for the recognition of ADL. Due to the limitation of number of sensors available in the off-the-shelf mobile devices, based on the previous study

[13] that uses only the accelerometer sensor, this study proposes the creation of two different methods for the recognition of ADL using different number of sensors in order to adapt the method according to the number of sensors available. Firstly, it proposes the fusion of the data acquired from the accelerometer, and the magnetometer sensors. Secondly, it proposes the fusion of the data acquired from the accelerometer, gyroscope, and magnetometer sensors. The ADL proposed for the recognition are running, walking, going upstairs, going downstairs, and standing, consisting this research on the analysis of the performance of three types of ANN, such as Multilayer Perception (MLP) with Backpropagation, Feedforward neural network with Backpropagation, and Deep Learning. A dataset used for this research is composed by the sensors' data acquired in several experiments with people aged between 16 and 60 years old, distinct lifestyles, and a mobile device in the front pocket of their pants, performing the proposed ADL. This research was conducted with the use of three Java libraries, such as Neuroph [14], Encog [15], and DeepLearning4j [16], and different datasets of features, in order to identify the best dataset of features and type of ANN for the recognition of ADL, verifying that the best accuracy for the recognition of ADL with the two different methods proposed was achieved with Deep Learning methods.

This paragraph concludes the section 1, and this paper is organized as follows: Section 2 summarizes the literature review for the use of data fusion techniques with accelerometer, gyroscope, and magnetometer sensors; Section 3 presents the methods used on each stage of the architecture proposed. The results obtained are presented in the Section 4, presenting the discussion about these results in the Section 5. The conclusions of this study are presented in the Section 6.

## 2. Related Work

Data fusion techniques may be used with the data acquired from motion and magnetic sensors available in the off-the-shelf mobile devices, *i.e.,* accelerometer, gyroscope, and magnetometer, in order to improve the reliability of the methods for the recognition of Activities of Daily Living (ADL) [2].

Following the main focus of this paper, the accelerometer, the gyroscope, and the magnetometer are used by the authors of [17] with the Random Forest classifier for the recognition of standing, going downstairs, going upstairs, sitting, walking, and running activities, using the variance, the mean, the frequency of the point with maximum amplitude, the energy of the extremum value, the value of the point with maximum amplitude, the mean of the extremum value, the period of the extremum value, the sum of the difference between extremum values, the maximum value around the midpoint, the minimum value around midpoint, and the mean of the vector around midpoint as features, reporting an average accuracy of 99.7%.

In addition, Shoaib *et al.* [18] presented a method that also uses the Global Positionning System (GPS) receiver, implementing Artificial Neural Networks (ANN), *i.e.,* Multi-Layer Perceptron (MLP), Support Vector Machine (SVM), Naïve Bayes, Logistic regression, decision tree, K-Nearest Neighbor (KNN), and rule based classifiers, extracting the mean and the standard deviation of the raw signal from the accelerometer, gyroscope, and magnetometer, and the distance from the GPS data, in order to recognize running, walking, standing, sitting, going downstairs, and going upstairs with a reported accuracy between 69% to 99%.

The authors of [19] also fused the accelerometer, gyroscope, and magnetometer data with the barometer data, extracting mean, mean of absolute values, median, variance, standard deviation, 75$^{th}$ percentile, inter-quartile range, average absolute difference, binned distribution, energy, Sub-band Energies, Sub-band Energy Ratios, Signal Magnitude Area (SMA), Zero-Crossing Rate, Number of Peaks, Absolute Value of short-time Fourier Transform, Power of short-time Fourier Transform, Power Spectral Centroid, Average of Continuous Wavelet Transform at various Approximation Levels, Frequency Domain Entropy, Frequency and Amplitude of the most 4 contributing Frequency Components obtained using spectral Fast Orthogonal Search, Cross-Correlation between Levelled Vertical and Horizontal Acceleration Components, and Ratio of Altitude Change to Number of Peaks, and implementing a decision tree, in order to recognize several ADL, such as going downstairs, going upstairs, standing, walking on an escalator, and taking an elevator with a reported accuracy between 80% to 90%.

The major part of studies in the literature only fuses the accelerometer, and gyroscope data, as the authors of [20] used the Random Forests (RF) variable importance is used for feature selection in order to recognize walking, going upstairs, going downstairs, sitting, standing, and laying activities with a Two-stage continuous Hidden Markov Model (HMM), reporting an accuracy of 91.76%. The Hidden Markov Model (HMM) was also implemented in [21], which also implemented the decision tree and Random Forest methods, with accelerometer and gyroscope data for the recognition of going downstairs, going upstairs, and walking, with variance, mean, standard deviation, maximum, minimum, median, interquartile range, skewness, Kurtosis, and spectrum peak position of the accelerometer and gyroscope data as features, reporting an accuracy of 93.8%.

In [22], the authors recognized walking, standing, running, going downstairs, going upstairs, and laying activities with accelerometer and gyroscope data, extracting the mean, the energy, the standard deviation, the correlation, and the entropy of the sensors' data, implementing the J48 decision tree, the logistic regression, the MLP, and the SVM methods with a reported accuracy between 89.3% and 100%.

The authors of [23] implemented the Signal Magnitude Vector (SMV) algorithm with a Threshold based algorithm for feature extraction in order to recognize some ADL, such as walking, standing, sitting, and running with a reported accuracy around 90%.

According to [24], the Gaussian mixture model (GMM) and the time series shapelets, applied to the accelerometer and gyroscope data, allow the recognition of sitting, standing, walking, and running activities with mean and standard deviation as features, reporting an accuracy of 88.64%. The authors of [25] also used the mean and standard deviation as features for the application of KNN and SVM methods, in order to recognize walking, resting, running, going downstairs, and going upstairs with a reported accuracy higher than 90%.

The standard deviation, maximum, minimum, correlation coefficients, interquartile range, mean, Dynamic time warping distance (DTW), Fast Fourier Transform (FFT) coefficients, and wavelet energy are extracted as features from accelerometer and gyroscope sensors, in order to recognize walking, jumping, running, going downstairs, and going upstairs with several methods, such as SVM, KNN, MLP, and Random Forest, reporting an accuracy between 84.97% and 90.65% [12]. The authors of [26] extracted the same features for the recognition of walking, going upstairs, going downstairs, jumping, and jogging activities, implementing KNN, Random Forests and SVM methods, reporting an accuracy of 95%.

The authors of [27] extracted the variance, mean, minimum and maximum along the Y axis of the accelerometer, and the variance and mean along the X axis of the gyroscope, and implemented the SVM method for the recognition of running, walking, going downstairs, going upstairs, standing, cycling and sitting, which reports an accuracy of 96%.

In [28], the authors extracted the skewness, mean, minimum, maximum, standard deviation, kurtosis, median, and interquartile range from the accelerometer and gyroscope data, implementing the MLP, SVM, Least Squares Method (LSM), and Naïve Bayes classifiers for the recognition of falling activities with a reported accuracy of 87.5%.

The SVM, Random Forest, J48 decision tree, Naïve Bayes, MLP, Rpart, JRip, Bagging, and KNN were implemented in [29] for the recognition of going downstairs, going upstairs, lying, standing, and walking with the mean and standard deviation along the X, Y and Z axis of the accelerometer and the gyroscope signal as features, reporting an accuracy higher than 90%.

The Root Mean Square (RMS), minimum, maximum, and zero crossing rate for X, Y, and Z axis were extracted from the accelerometer and gyroscope data, and the ANOVA method was applied for the correct recognition of sitting, resting, turning, and walking with a reported accuracy around 100% [30].

The driving, walking, running, cycling, resting, and jogging was recognized by ANN with mean, minimum, maximum, standard deviation, difference between maximum and minimum, Parseval's Energy, Parseval's Energy in the frequency range 0 - 2.5 Hz, Parseval's Energy in the frequencies greater than 2.5 Hz, RMS, kurtosis, correlation between axis, ratio of the maximum and minimum values in the FFT, skewness, difference between the maximum and minimum values in the FFT, median of troughs, median of peaks, number of troughs, number of peaks, average distance between two consecutive troughs, average distance between two consecutive peaks, indices of the 8 highest peaks after the application of the FFT, and ratio of the average values of peaks and troughs as features from the accelerometer and gyroscope data, reporting an accuracy between 57.53% to 97.58% [31].

The Threshold Based Algorithm (TBA) was applied to the values of the acceleration, and the difference between adjacent elements of the heading, extracted from the accelerometer and gyroscope sensors, in order to recognize going downstairs, going upstairs, running, walking, and jumping with a reported accuracy of 83% [32].

The median absolute deviation, minimum, maximum, absolute mean, interquartile range, Signal Magnitude Range, skewness, and Kurtosis were extracted from accelerometer and gyroscope signal for the application of KNN, SVM, Sparse Representation Classifier, and Kernel-Extreme Learning Machine, in order to recognize standing, running, going upstairs, walking, and going downstairs, reporting an average accuracy of 94.5% [33].

The jogging and walking activities are recognized with mean, variance, minimum, and maximum of the X, Y and Z axis of the accelerometer and gyroscope sensors as features applied to the SVM method, reporting an accuracy of 95.5% [34].

The authors of [35] implemented sparse approximation, KNN, SVM, Spearman correlation, Fuzzy c-means, MLP, and linear regression classifiers for the recognition of running, cycling, sitting, walking, and standing, using the standard deviation, mean, median, power ratio of the frequency bands, peak acceleration, and energy extracted from the accelerometer and gyroscope signal, reporting an accuracy of 98%.

In [36], the implementation of SVM and Random Forest methods was used for the recognition of standing, sitting, laying, walking, going downstairs, and going upstairs, with the extraction of the angle, the minimum, the maximum, and the mean values of the accelerometer and gyroscope signal, reporting an accuracy around 100%.

The authors of [37] used the accelerometer and gyroscope sensors for the recognition of the movements related to up and down buses, implementing the C4.5 decision tree, Naïve Bayes, KNN, logistic regression, SVM, and MLP with mean, standard deviation, energy, correlation between axis, and magnitude of FFT components as features, reporting an accuracy of 95.3%.

The accelerometer, gyroscope, barometer, and GPS were used for the recognition of standing, sitting, washing dishes, going downstairs, going upstairs, walking, running, and cycling with standard deviation, mean, interquartile range, mean squared, altitude difference in meters, and speed as features applied to the SVM method, whose the authors reported an accuracy around 90% [38].

For the recognition of walking, lying, running, cycling, jogging, washing dishes, vacuum cleaning, playing piano, playing cello, playing tennis, brushing teeth, wiping cupboard, driving, taking an elevator, doing laundry, working on a computer, eating, reading a book, going downstairs, going upstairs, and folding laundry, the authors of [39] used the features extracted from the accelerometer, the gyroscope, and the camera, including the variance and mean for each axis, the movement intensity, the energy, the energy consumption, and the periodicity, applying them to the HMM, the SVM, the Naïve Bayes methods, and obtaining a reported accuracy of 81.5%.

The J48 decision tree, IBk, MLP, and Logistic regression methods were implemented with the median, the mean, the standard deviation, the kurtosis, the skewness, the maximum, the minimum, the slope, difference between maximum and minimum, the spectral centroid, the entropy of the energy in 10 equal sized blocks, the short time energy, the spectral roll off, the zero crossing rate, the spectral flux, and the spectral centroid for each axis and the absolute value of the accelerometer, gyroscope, and orientation sensors [40], in order to recognize walking, standing, jogging, going downstairs, going upstairs, jumping, and sitting activities, reporting an accuracy of 94%.

According to the analysis previously presented, table 1 shows the ADL recognized with the use of the accelerometer, gyroscope and/or magnetometer sensors, verifying that the walking, standing/resting, going downstairs, going upstairs, running, and sitting are the most recognized ADL. The lines in this table are sorted in decreasing manner regarding the number of studies found for each activity. Shown in a darker background, the activities reported in at least 10 papers.

*Table 1 - Distribution of the ADL extracted in the studies analyzed.*

| ADL: | Number of Studies: |
|---|---|
| walking | 21 |
| going downstairs | 17 |
| going upstairs | 17 |
| standing/resting | 16 |
| running | 13 |
| sitting | 11 |
| laying | 5 |
| jogging | 5 |
| cycling | 5 |
| jumping | 4 |
| taking an elevator | 2 |
| driving | 2 |
| washing dishes | 2 |

| ADL: | Number of Studies: |
|---|---|
| walking on an escalator | 1 |
| turning | 1 |
| vacuum cleaning | 1 |
| playing piano | 1 |
| playing cello | 1 |
| playing tennis | 1 |
| brushing teeth | 1 |
| wiping cupboard | 1 |
| doing laundry | 1 |
| folding laundry | 1 |
| working on a computer | 1 |
| reading a book | 1 |
| eating | 1 |

The features used in the recognition of the ADL are presented in table 2, showing that the mean, standard deviation, maximum, minimum, energy, inter-quartile range, correlation coefficients, median, and variance are the most used features, with more relevance for mean, standard deviation, maximum, and minimum. This table is sorted in decreasing order of the number of studies that reportedly used a specific feature. In darker background the features used in 6 or more papers.

*Table 2 - Distribution of the features extracted in the studies analyzed.*

| Features: | Number of Studies: |
|---|---|
| mean | 20 |
| standard deviation | 15 |
| maximum | 12 |
| minimum | 12 |
| energy | 10 |
| inter-quartile range | 7 |
| variance | 6 |
| median | 6 |
| correlation coefficients | 6 |
| skewness | 5 |
| Kurtosis | 5 |
| Zero-Crossing Rate | 3 |
| Power Spectral Centroid | 3 |
| Frequency Domain Entropy | 3 |
| period of the extremum value | 2 |
| mean of absolute values | 2 |
| Signal Magnitude Area (SMA) | 2 |
| Dynamic time warping distance (DTW) | 2 |
| Fast Fourier Transform (FFT) coefficients | 2 |
| Root Mean Square (RMS) | 2 |
| difference between maximum and minimum | 2 |
| frequency of the point with maximum amplitude | 1 |
| energy of the extremum value | 1 |
| value of the point with maximum amplitude | 1 |
| mean of the extremum value | 1 |
| sum of the difference between extremum values | 1 |
| mean of the vector around midpoint | 1 |
| 75th percentile | 1 |
| average absolute difference | 1 |
| Sub-band Energies | 1 |
| Sub-band Energy Ratios | 1 |

| Features: | Number of Studies: |
|---|---|
| Number of Peaks | 1 |
| Absolute Value of short-time Fourier Transform | 1 |
| Power of short-time Fourier Transform | 1 |
| Average of Continuous Wavelet Transform at various Approximation Levels | 1 |
| Frequency and Amplitude of the most 4 contributing Frequency Components obtained using spectral Fast Orthogonal Search | 1 |
| Ratio of Altitude Change to Number of Peaks | 1 |
| spectrum peak position | 1 |
| ratio of the maximum and minimum values in the FFT | 1 |
| difference between the maximum and minimum values in the FFT | 1 |
| median of troughs | 1 |
| median of peaks | 1 |
| number of troughs | 1 |
| number of peaks | 1 |
| average distance between two consecutive troughs | 1 |
| average distance between two consecutive peaks | 1 |
| slope | 1 |

Finally, the methods implemented for the recognition of the ADL in the literature are presented in table 3, concluding that the methods with an accuracy higher than 90% are MLP, logistic regression, random forest and decision tree methods, verifying that the method that reports the best average accuracy in the recognition of ADL is the MLP, with an average accuracy equals to 93.86%.

*Table 3 - Distribution of the classification methods used in the studies analyzed.*

| Methods: | Number of Studies: | Average of Reported Accuracy: |
|---|---|---|
| Artificial Neural Networks (ANN) / Multi-Layer Perceptron (MLP) | 9 | 93.86% |
| Logistic regression | 4 | 92.18% |
| Random Forest | 6 | 90% |
| Decision trees (J48, C4.5) | 3 | 90.89% |
| ANOVA | 1 | 88% |
| Support Vector Machine (SVM) | 15 | 88.1% |
| K-Nearest Neighbor (KNN) | 8 | 85.67% |
| Hidden Markov Model (HMM) | 3 | 84.22% |
| Threshold Based Algorithm (TBA) | 1 | 83% |
| Naïve Bayes | 6 | 82.86% |
| Least Squares Method (LSM) | 1 | 80% |
| rule based classifiers (J-Rip, Rpart) | 3 | 76.35% |
| IBk | 1 | 76.28% |
| Sparse Representation Classifier | 2 | - |
| binned distribution | 1 | - |
| Signal Magnitude Vector (SMV) | 1 | - |
| Gaussian mixture model (GMM) | 1 | - |
| time series shapelets | 1 | - |
| Bagging | 1 | - |
| Kernel-Extreme Learning Machine | 1 | - |
| Spearman correlation | 1 | - |
| Fuzzy c-means | 1 | - |
| Linear Regression | 1 | - |

## 3. Methods

Based on the related work presented in the previous section and the proposed architecture of a framework for the recognition of ADL previously presented in [4-6], there are several modules for the creation of the final method, such as data acquisition, data processing, data fusion, and artificial intelligence methods. Assuming that the data acquired from all sensors is fulfilled, the data processing module is composed by data cleaning and feature extraction methods. Data fusion and artificial intelligence methods are commonly performed in parallel.

Section 3.1 presents the methodology for the data acquisition. Data processing methods are presented in the section 3.2. And, finally, in the section 3.3, the data fusion and artificial intelligence method are presented.

### 3.1. Data Acquisition

A mobile application developed for Android devices [41, 42] was installed in a BQ Aquarius device [43] for the acquisition of the sensors' data, saving the data captured from the accelerometer, magnetometer, and gyroscope sensors in text files. The mobile application captures the data in 5 seconds slots every 5 minutes, where the frequency of the data acquisition is around 10ms. For the definition of the experiments, 25 individuals aged between 16 and 60 years old were selected. The individuals selected had distinct lifestyles, where 10 individuals are active and the remaining 15 individuals are mainly sedentary. During the data acquisition, the mobile device should be in the pocket of the user, and the user should define a label of the ADL performed in each 5 seconds of data captured. Based on the ADL that are the most identified in the previous research studies, the selected ADL for this study in the mobile application are running, walking, going upstairs, going downstairs, and standing. After the data acquisition process, a set of 2000 captures for each ADL, each with 5 seconds of raw data, was stored in the ALLab MediaWiki [44].

### 3.2. Data Processing

The data processing is the second step of the method for the recognition of ADL, which is executed after the data acquisition. Data cleaning methods are executed for the noise reduction, as presented in section 3.2.1. After cleaning the data, the features were extracted for further analysis, as discussed in section 3.2.2.

#### 3.2.1. Data Cleaning

Data cleaning is a process to filter the data acquired from the accelerometer, magnetometer, and gyroscope sensors, in order to remove the noise. The data cleaning method should be selected according to the types of sensors used, but the low-pass filter is the best method for the data acquired from the sensors used in this study [45], removing the noise and allowing the correct extraction of the selected features.

#### 3.2.2. Feature Extraction

After the data cleaning and based on the features most commonly extracted in previous research studies (see Table 2), several features were extracted from the accelerometer, magnetometer, and gyroscope sensors. These are the five greatest distances between the maximum peaks, the Average of the maximum peaks, the Standard Deviation of the maximum peaks, the Variance of the maximum peaks, the Median of the maximum peaks, the Standard Deviation of the raw signal, the Average of the raw signal, the Maximum value of the raw signal, the Minimum value of the raw signal, the Variance of the of the raw signal, and the Median of the raw signal.

### 3.3. Identification of Activities of Daily Living with Data Fusion

Extending a previous study [13] that used only the accelerometer sensor, this study fuses the features extracted from the accelerometer and magnetometer sensors (section 3.3.1), and the features extracted from the accelerometer, gyroscope and magnetometer sensors (section 3.3.2). Finally, the artificial intelligence methods for the identification of ADL are presented in the section 3.3.3.

### 3.3.1. Data Fusion with Accelerometer and Magnetometer sensors

Regarding the features extracted from each ADL, five datasets have been constructed with features extracted from the accelerometer and magnetometer sensors' data acquired during the performance of the five ADL, having 2000 records from each ADL. The datasets defined are:

- **Dataset 1:** Composed by the five greatest distances between the maximum peaks, Average of the maximum peaks, Standard Deviation of the maximum peaks, Variance of the maximum peaks, Median of the maximum peaks, Standard Deviation of the raw signal, Average of the raw signal, Maximum value of the raw signal, Minimum value of the raw signal, Variance of the of the raw signal, and Median of the raw signal, extracted from the accelerometer and the magnetometer sensors;
- **Dataset 2:** Composed by Average of the maximum peaks, Standard Deviation of the maximum peaks, Variance of the maximum peaks, Median of the maximum peaks, Standard Deviation of the raw signal, Average of the raw signal, Maximum value of the raw signal, Minimum value of the raw signal, Variance of the of the raw signal, and Median of the raw signal, extracted from the accelerometer and the magnetometer sensors;
- **Dataset 3:** Composed by Standard Deviation of the raw signal, Average of the raw signal, Maximum value of the raw signal, Minimum value of the raw signal, Variance of the of the raw signal, and Median of the raw signal, extracted from the accelerometer and the magnetometer sensors;
- **Dataset 4:** Composed by Standard Deviation of the raw signal, Average of the raw signal, Variance of the of the raw signal, and Median of the raw signal, extracted from the accelerometer and the magnetometer sensors;
- **Dataset 5:** Composed by Standard Deviation of the raw signal, and Average of the raw signal, extracted from the accelerometer and the magnetometer sensors.

### 3.3.2. Data Fusion with Accelerometer, Magnetometer and Gyroscope sensors

Regarding the features extracted from each ADL, five datasets have been constructed with features extracted from the accelerometer, magnetometer and gyroscope sensors' data acquired during the performance of the five ADL, having 2000 records from each ADL. The datasets defined are:

- **Dataset 1:** Composed by the five greatest distances between the maximum peaks, Average of the maximum peaks, Standard Deviation of the maximum peaks, Variance of the maximum peaks, Median of the maximum peaks, Standard Deviation of the raw signal, Average of the raw signal, Maximum value of the raw signal, Minimum value of the raw signal, Variance of the of the raw signal, and Median of the raw signal, extracted from the accelerometer, the magnetometer and the gyroscope sensors;
- **Dataset 2:** Composed by Average of the maximum peaks, Standard Deviation of the maximum peaks, Variance of the maximum peaks, Median of the maximum peaks, Standard Deviation of the raw signal, Average of the raw signal, Maximum value of the raw signal, Minimum value of the raw signal, Variance of the of the raw signal, and Median of the raw signal, extracted from the accelerometer, the magnetometer and the gyroscope sensors;
- **Dataset 3:** Composed by Standard Deviation of the raw signal, Average of the raw signal, Maximum value of the raw signal, Minimum value of the raw signal, Variance of the of the raw signal, and Median of the raw signal, extracted from the accelerometer, the magnetometer and the gyroscope sensors;
- **Dataset 4:** Composed by Standard Deviation of the raw signal, Average of the raw signal, Variance of the of the raw signal, and Median of the raw signal, extracted from the accelerometer, the magnetometer and the gyroscope sensors;
- **Dataset 5:** Composed by Standard Deviation of the raw signal, and Average of the raw signal, extracted from the accelerometer, the magnetometer and the gyroscope sensors.

### 3.3.3. Artificial Intelligence

Based on the results reported by the literature review presented in the section 2, one of the most used methods for the recognition of ADL based on the use of the mobiles' sensors is the ANN, and this method reports a better accuracy than SVM, KNN, Random Forest, and Naïve Bayes.

Following the datasets defined in the sections 3.3.1 and 3.3.2, this study implements three types of neural networks, such as MLP, Feedforward Neural Network, and Deep Neural Networks (DNN), in order to identify the best neural network for the recognition of ADL, these are:

- MLP with Backpropagation, applied with Neuroph framework [14];
- Feedforward Neural Network with Backpropagation, applied with Encog framework [15];
- Deep Neural Networks, applied with DeepLearning4j framework [16].

In order to improve the results obtained by the neural networks, the MIN/MAX normalizer [46] was applied to the defined datasets, implementing the MLP with Backpropagation, and the Feedforward Neural Network with Backpropagation with normalized and non-normalized at different stages.

Before the application of the Deep Neural Networks, the $L_2$ regularization [47] was applied to the defined datasets. After the application of the $L_2$ regularization, the normalization with mean and standard deviation [48] was applied to the datasets, implementing Deep Neural Networks method with normalized and non-normalized data at different stages.

The number of training iterations may influence the results of the neural networks, defining the maximum number of $10^6$, $2 \times 10^6$ and $4 \times 10^6$ iterations, in order to identify the best number of training iterations with best results.

After this research, the methods that should be implemented in the framework for the recognition of ADL defined in [4-6] is a function of the number of sensors available in the off-the-shelf mobile device. According to the results available in [13], if the mobile device has only the accelerometer sensor, the method that should be implemented is the Deep Neural Networks, verifying with this research the best methods for the use of the datasets defined in the sections 3.3.1 and 3.3.2.

### 4. Results

This research consists in the creation of two different methods for the recognition of ADL with different number of sensors. Firstly, the results of the creation of a method with accelerometer and magnetometer sensors are presented in the section 4.1. Finally, the results of the creation of a method with accelerometer, magnetometer, and gyroscope sensors are presented in the section 4.2.

### 4.1. Identification of Activities of Daily Living with Accelerometer and Magnetometer sensors

Based on the datasets defined in the section 3.3.1, the three types of neural networks proposed in the section 3.3.3 are implemented with the frameworks proposed, these are MLP with Backpropagation, Feedforward Neural Network with Backpropagation, and Deep Neural Networks. The defined training dataset has 10000 records, where each ADL has 2000 records.

Firstly, the results of the implementation of the MLP with Backpropagation using the Neuroph framework are presented in the figure 1, verifying that the results have very low accuracy with all datasets, achieving values between 20% and 40% with non-normalized data (figure 1-a), and values between 20% and 30% with normalized data (figure 1-b).

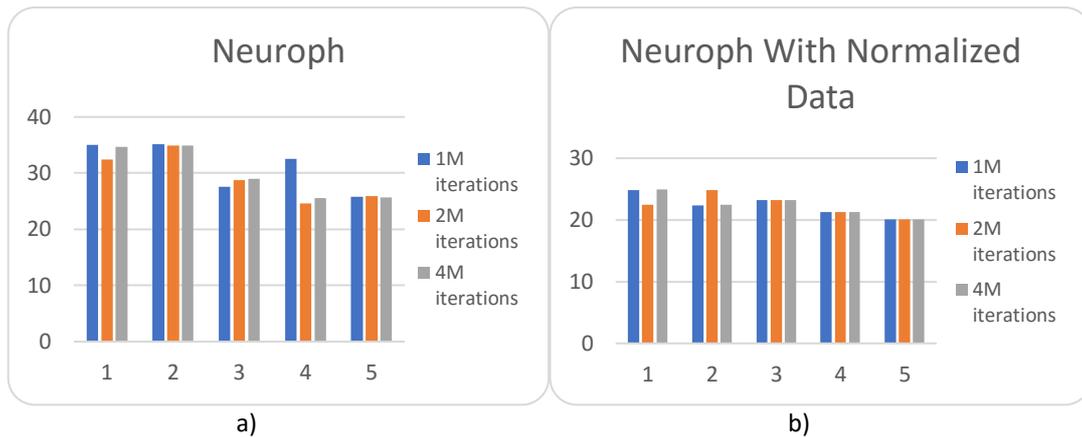

*Figure 1 –Results obtained with Neuroph framework for the different datasets of accelerometer and magnetometer sensors (horizontal axis) and different maximum number of iterations (series), obtaining the accuracy in percentage (vertical axis). The figure a) shows the results with data without normalization. The figure b) shows the results with normalized data.*

Secondly, the results of the implementation of the Feedforward Neural Network with Backpropagation using the Encog framework are presented in the figure 2. In general, this type of neural network achieves bad results with both non-normalized and normalized data, reporting the maximum results around 40%. With non-normalized data (figure 2-a), the neural networks reports results above 30% with the dataset 1 trained over 1M and 4M iterations, the dataset 2 trained over 1M iterations, the dataset 3 trained over 2M iterations, the dataset 4 trained over 1M, 2M and 4M iterations, and the dataset 5 trained over 1M and 4M iterations. With normalized data (figure 2-b), the results reported are lower than 40%, with an exception for the neural network trained over 2M iterations with the dataset 5 that reports results higher than 60%.

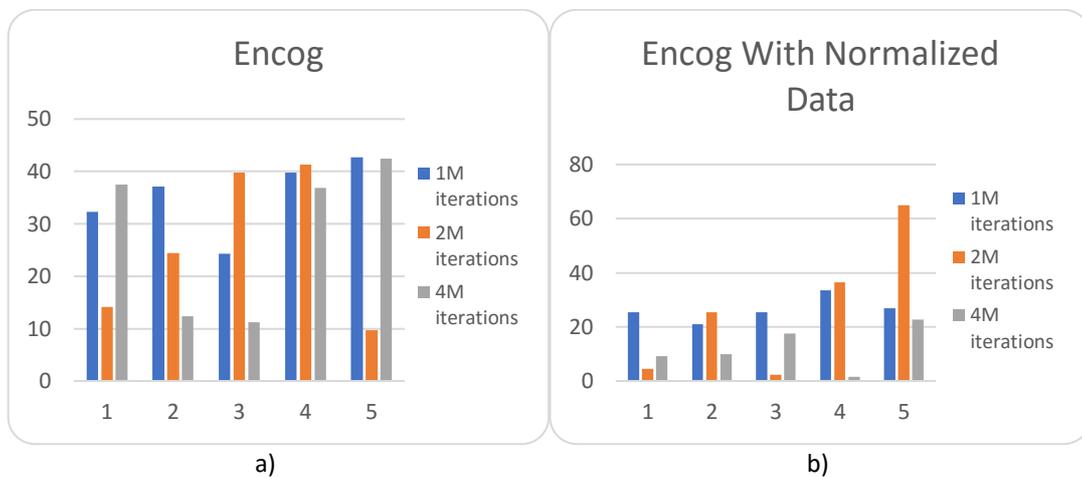

*Figure 2 –Results obtained with Encog framework for the different datasets of accelerometer and magnetometer sensors (horizontal axis) and different maximum number of iterations (series), obtaining the accuracy in percentage (vertical axis). The figure a) shows the results with data without normalization. The figure b) shows the results with normalized data.*

Finally, the results of the implementation of Deep Neural Networks with DeepLearning4j framework are presented in the figure 3. With non-normalized data (figure 3-a), the results obtained are below the expectations (around 20%) for the datasets 2, 3 and 4, and the results obtained with dataset 5 are around 70%. On the other hand, with normalized data (figure 3-b), the results reported are always higher than 70%, achieving better results with the dataset 1, decreasing with the reduction of the number of features in the dataset.

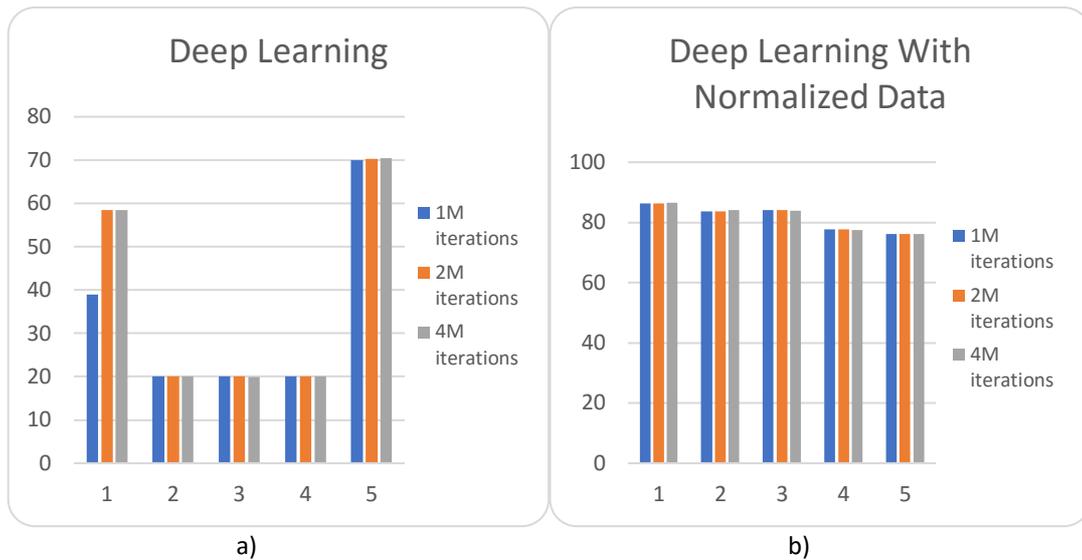

*Figure 3 – Results obtained with DeepLearning4j framework for the different datasets of accelerometer and magnetometer sensors (horizontal axis) and different maximum number of iterations (series), obtaining the accuracy in percentage (vertical axis). The figure a) shows the results with data without normalization. The figure b) shows the results with normalized data.*

In table 4, the maximum accuracies achieved with the different types of neural networks are presented with the relation of the different datasets used for accelerometer and magnetometer data, and the maximum number of iterations, verifying that the use of Deep Neural Networks with normalized data reports better results than others.

*Table 4 - Best accuracies obtained with the different frameworks, datasets and number of iterations.*

|  | FRAMEWORK | DATASET | ITERATIONS NEEDED FOR TRAINING | BEST ACCURACY ACHIEVED (%) |
|---|---|---|---|---|
| NOT NORMALIZED DATA | NEUROPH | 2 | 1M | 35.15 |
|  | ENCOG | 5 | 1M | 42.75 |
|  | DEEP LEARNING | 5 | 4M | 70.43 |
| NORMALIZED DATA | NEUROPH | 1 | 4M | 24.93 |
|  | ENCOG | 5 | 2M | 64.94 |
|  | DEEP LEARNING | 1 | 4M | 86.49 |

Regarding the results obtained, in the case of the use of accelerometer and magnetometer sensors in the framework for the identification of ADL, the type of neural networks that should be used is a Deep Neural Network (Deep Learning) with normalized data, because the results obtained are always higher than 80%.

### 4.2. Identification of Activities of Daily Living with Accelerometer, Magnetometer and Gyroscope sensors

Based on the datasets defined in the section 3.3.2, the three types of neural networks proposed in the section 3.3.3 are implemented with the frameworks proposed, these are MLP with Backpropagation, Feedforward Neural Network with Backpropagation, and Deep Neural Networks. The defined training dataset has 10000 records, where each ADL has 2000 records.

Firstly, the results of the implementation of the MLP with Backpropagation using the Neuroph framework are presented in the figure 4, verifying that the results have very low accuracy with all datasets. With non-normalized data (figure 4-a), the results achieved are between 20% and 40%, where the better accuracy was achieved with the dataset 2. And, with normalized data (figure 4-b), the results obtained are between 30% and 40%, with lower results with the dataset 5.

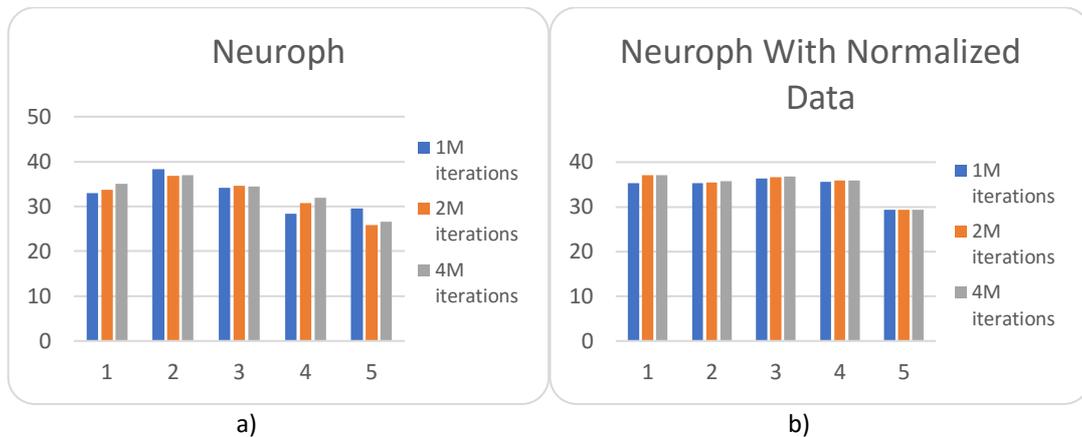

*Figure 4 –Results obtained with Neuroph framework for the different datasets of accelerometer, magnetometer and gyroscope sensors (horizontal axis) and different maximum number of iterations (series), obtaining the accuracy in percentage (vertical axis). The figure a) shows the results with data without normalization. The figure b) shows the results with normalized data.*

Secondly, the results of the implementation of the Feedforward Neural Network with Backpropagation using the Encog framework are presented in the figure 5. In general, this type of neural network achieves bad results with non-normalized and normalized data, reporting the maximum results around 40%. With non-normalized data (figure 5-a), the neural network reports results above 30% with the dataset 2 trained over 2M iterations, the dataset 3 trained over 4M iterations, and the dataset 4 trained over 4M iterations, reporting an accuracy higher than 70% with the dataset 2 trained over 2M iterations. With normalized data (figure 5-b), the results reported are lower than 20%, with an exception for the neural network with the dataset 3 trained over 4M iterations, the dataset 4 trained over 2M iterations, and the dataset 5 trained over 1M and 2M iterations.

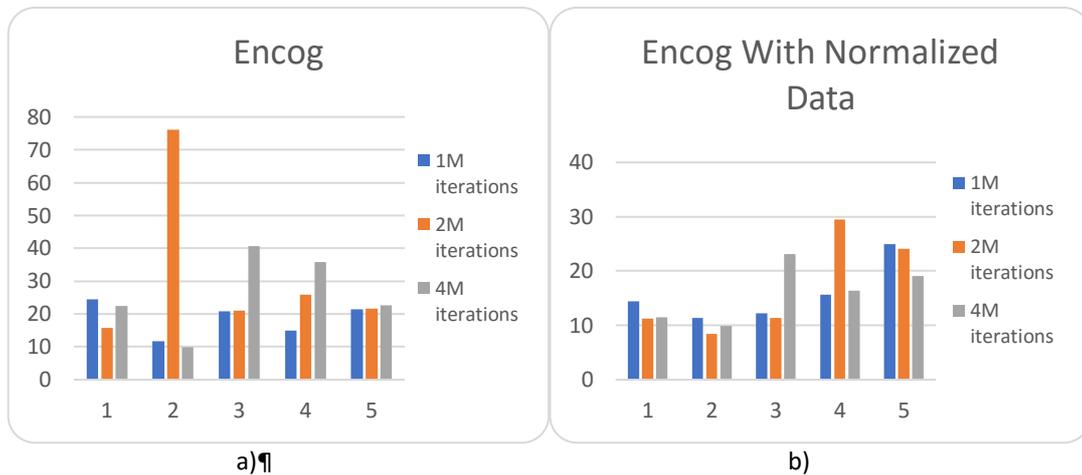

*Figure 5 –Results obtained with Encog framework for the different datasets of accelerometer, magnetometer and gyroscope sensors (horizontal axis) and different maximum number of iterations (series), obtaining the accuracy in percentage (vertical axis). The figure a) shows the results with data without normalization. The figure b) shows the results with normalized data.*

Finally, the results of the implementation of Deep Neural Networks with DeepLearning4j framework are presented in the figure 6. With non-normalized data (figure 6-a), the results obtained are below the expectations (around 40%) for the datasets 2, 3 and 4, and the results obtained with datasets 1 and 5 are around 70%. On the other hand, with normalized data (figure 6-b), the results reported are always higher than 80%, achieving better results with the dataset 1, decreasing with the reduction of the number of features in the dataset.

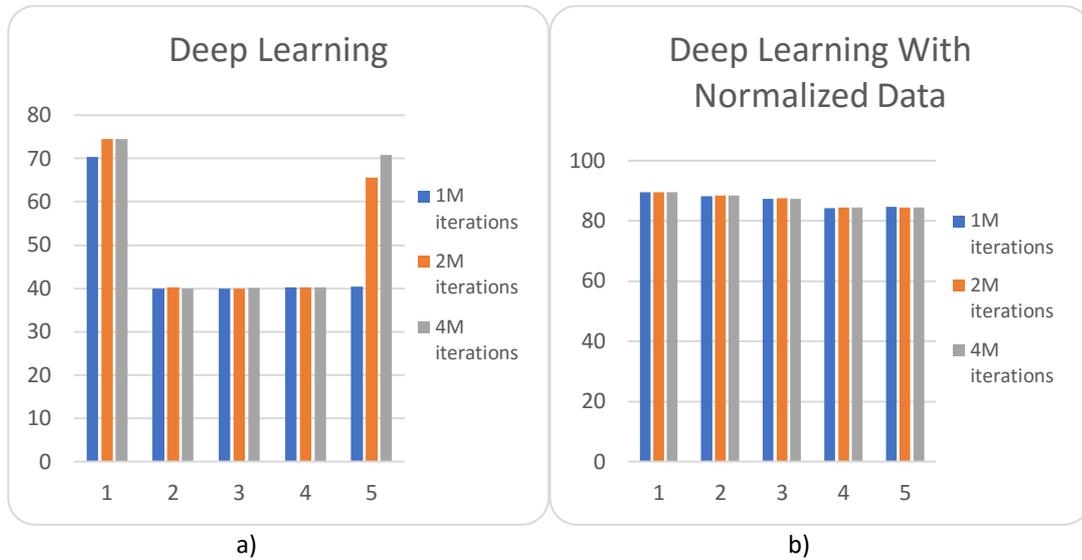

*Figure 6 –Results obtained with DeepLearning4j framework for the different datasets of accelerometer, magnetometer and gyroscope sensors (horizontal axis) and different maximum number of iterations (series), obtaining the accuracy in percentage (vertical axis). The figure a) shows the results with data without normalization. The figure b) shows the results with normalized data.*

In table 5, the maximum accuracies achieved with the different types of neural networks are presented with the relation of the different datasets used for the accelerometer, magnetometer and gyroscope data, and the maximum number of iterations, verifying that the use of Deep Neural Networks with normalized data reports better results than others.

*Table 5 - Best accuracies obtained with the different frameworks, datasets and number of iterations.*

| | FRAMEWORK | DATASETS | ITERATIONS NEEDED FOR TRAINING | BEST ACCURACY ACHIEVED (%) |
|---|---|---|---|---|
| **NOT NORMALIZED DATA** | NEUROPH | 2 | 1M | 38.32 |
| | ENCOG | 2 | 2M | 76.13 |
| | DEEP LEARNING | 1 | 2M | 74.47 |
| **NORMALIZED DATA** | NEUROPH | 1 | 2M | 37.13 |
| | ENCOG | 4 | 2M | 29.54 |
| | DEEP LEARNING | 1 | 4M | 89.51 |

Regarding the results obtained, in the case of the use of accelerometer, magnetometer and gyroscope sensors in the framework for the identification of ADL, the type of neural networks that should be used is a Deep Neural Network (Deep Learning) with normalized data, because the results obtained are always higher than 80%, and the best result was achieved with dataset 1 that was equals to 89.51%.

## 5. Discussion

Following the research for the development of a framework for the identification of the ADL using motion and magnetic sensors, presented in [4-6], there are several modules, such as data acquisition, data cleaning, feature extraction, data fusion, and artificial intelligence methods. The choice of the methods for data fusion, and artificial intelligence modules, depends on the number of sensors available on the mobile device, using the maximum number of sensors available on the mobile device, in order to increase the reliability of the method. In the figure 7, a simplified schema for the development of a framework for the identification of ADL is presented.

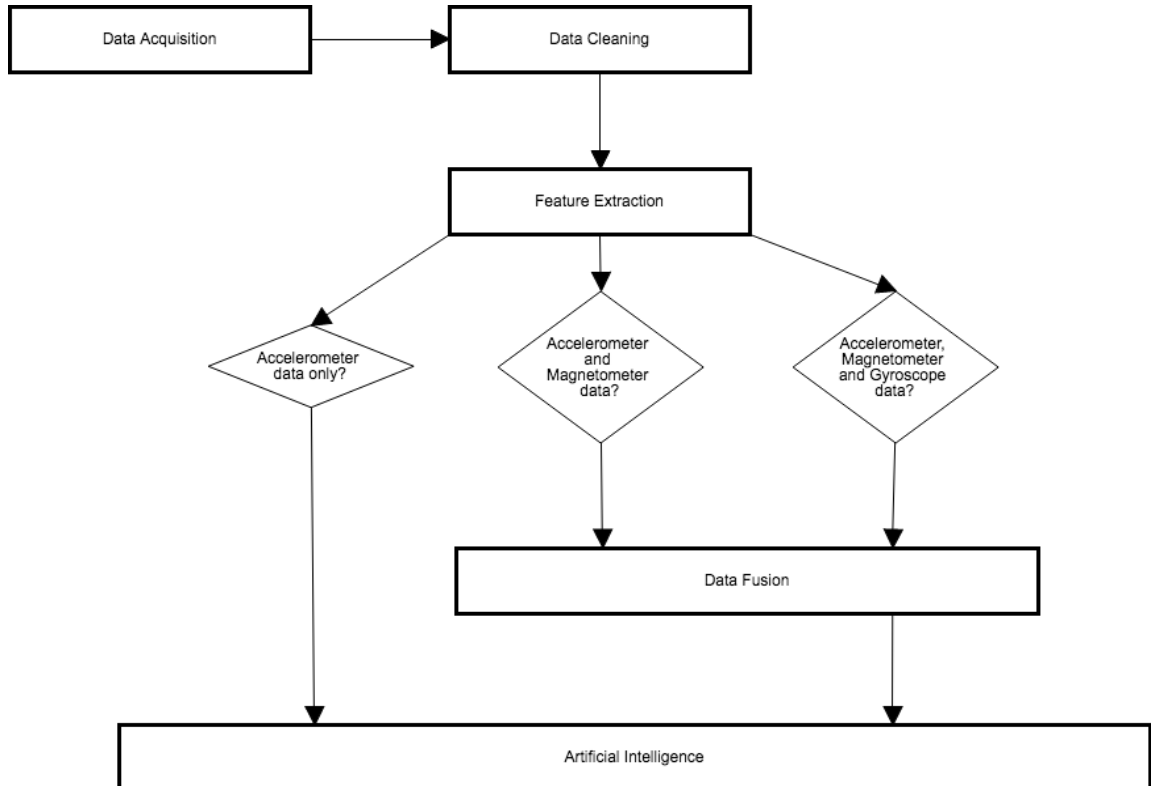

*Figure 7 - Simplified diagram for the framework for the identification of ADL.*

According to the previous study based only in the use of the accelerometer sensor for the recognition of ADL, presented in [13], the best results achieved for each type of neural network are presented in the table 6, verifying that the best method is Deep Neural Networks with normalized data, reporting an accuracy of 85.89%. In the case of the mobile device only has the accelerometer sensor available, Deep Neural Networks with normalized data should be implemented in the framework for the recognition of ADL, removing the data fusion, as presented in the figure 7.

*Table 6 –Best accuracies achieved by the method using only the accelerometer sensor.*

|  | FRAMEWORK | BEST ACCURACY ACHIEVED (%) |
|---|---|---|
| NOT NORMALIZED DATA | NEUROPH | 34.76 |
|  | ENCOG | 74.45 |
|  | DEEP LEARNING | 80.35 |
| NORMALIZED DATA | NEUROPH | 24.03 |
|  | ENCOG | 37.07 |
|  | DEEP LEARNING | 85.89 |

Based on results obtained with the use of accelerometer and magnetometer sensors, presented in the section 4.1, the comparison of the results between the use of the accelerometer sensor, and the use of accelerometer and magnetometer sensors is presented in the table 7. In general, the accuracy increases with the use of normalized data, and decreases with the use of non-normalized data, where the highest difference was verified with the use of the accelerometer and magnetometer sensors with the implementation of Feedforward Neural Network with Backpropagation using the Encog framework, reporting a difference of 27.87%. However, the Deep Neural Networks (Deep Learning) continues achieving the better results with an accuracy of 86.49%. In the case of the mobile device only has the accelerometer and magnetometer sensors available, Deep Neural Networks with normalized data should be implemented in the framework for the recognition of ADL, as presented in the figure 7.

Table 7 - Comparison between the best results achieved only using the accelerometer sensor, and using the accelerometer and magnetometer sensors.

|  | FRAMEWORK | BEST ACCURACY ACHIEVED (%) | | DIFFERENCE (%) |
|---|---|---|---|---|
|  |  | ACCELEROMETER | ACCELEROMETER MAGNETOMETER |  |
| NOT NORMALIZED DATA | NEUROPH | 34.76 | 35.15 | +0.39 |
| NOT NORMALIZED DATA | ENCOG | 74.45 | 42.75 | -31.70 |
| NOT NORMALIZED DATA | DEEP LEARNING | 80.35 | 70.43 | -9.92 |
| NORMALIZED DATA | NEUROPH | 24.03 | 24.93 | +0.90 |
| NORMALIZED DATA | ENCOG | 37.07 | 64.94 | +27.87 |
| NORMALIZED DATA | DEEP LEARNING | 85.89 | 86.49 | +0.60 |

Based on results obtained with the use of accelerometer, magnetometer and gyroscope sensors, presented in the section 4.2, the comparison of the results between the use of the accelerometer sensor, and the use of accelerometer, magnetometer and gyroscope sensors is presented in the table 8. In general, the accuracy increases, except in the cases of the use of Deep Neural Networks (Deep Learning) with non-normalized data and Feedforward Neural Network with Backpropagation using the Encog framework with normalized data. The highest difference in the accuracy is verified with the use of MLP with Backpropagation using the Neuroph framework, where the accuracy results increased 13.1% with normalized data, but the Deep Neural Networks (Deep Learning) achieves better results with an accuracy of 89.51%.

Table 8 - Comparison between the best results achieved only using the accelerometer sensor, and using the accelerometer, magnetometer and gyroscope sensors.

|  | FRAMEWORK | BEST ACCURACY ACHIEVED (%) | | DIFFERENCE (%) |
|---|---|---|---|---|
|  |  | ACCELEROMETER | ACCELEROMETER MAGNETOMETER GYROSCOPE |  |
| NOT NORMALIZED DATA | NEUROPH | 34.76 | 38.32 | +3.56 |
| NOT NORMALIZED DATA | ENCOG | 74.45 | 76.13 | +1.46 |
| NOT NORMALIZED DATA | DEEP LEARNING | 80.35 | 74.47 | -5.88 |
| NORMALIZED DATA | NEUROPH | 24.03 | 37.13 | +13.10 |
| NORMALIZED DATA | ENCOG | 37.07 | 29.54 | -7.53 |
| NORMALIZED DATA | DEEP LEARNING | 85.89 | 89.51 | +3.62 |

Based on results obtained with the use of accelerometer, magnetometer and gyroscope sensors, presented in the section 4.2, and the results obtained with the use of accelerometer and magnetometer sensors, presented in the section 4.1, the comparison between these results is presented in the table 9. In general, the accuracy increases, except in the case of the use of Feedforward Neural Network with

Backpropagation using the Encog framework with normalized data. The highest different in the accuracy is verified with the use of Feedforward Neural Network with Backpropagation using the Encog framework with non-normalized data, where the accuracy results increased 33.38% with non-normalized data, but the Deep Neural Networks (Deep Learning) continues achieving the better results with an accuracy of 89.51%. Thus, in the case of the mobile device has the accelerometer, magnetometer and gyroscope sensors available, Deep Neural Networks with normalized data should be implemented in the framework for the recognition of ADL, as presented in the figure 7.

*Table 9 - Comparison between the best results achieved only using the accelerometer and magnetometer sensors, and using the accelerometer, magnetometer and gyroscope sensors*

|  | FRAMEWORK | BEST ACCURACY ACHIEVED (%) | | DIFFERENCE (%) |
|---|---|---|---|---|
|  |  | ACCELEROMETER MAGNETOMETER | ACCELEROMETER MAGNETOMETER GYROSCOPE |  |
| NOT NORMALIZED DATA | NEUROPH | 35.15 | 38.32 | +3.17 |
|  | ENCOG | 42.75 | 76.13 | +33.38 |
|  | DEEP LEARNING | 70.43 | 74.47 | +4.04 |
| NORMALIZED DATA | NEUROPH | 24.93 | 37.13 | +12.20 |
|  | ENCOG | 64.94 | 29.54 | -35.40 |
|  | DEEP LEARNING | 86.49 | 89.51 | +3.02 |

In conclusion, when compared with MLP with Backpropagation using the Neuroph framework and Feedforward Neural Network with Backpropagation using the Encog framework, the Deep Neural Networks with normalized data achieves better results for the recognition of the ADL with accuracies between 85% and 90%.

## 6. Conclusions

The sensors that are available in the mobile devices, including accelerometer, gyroscope, and magnetometer sensors, allow the capture of data that can be used to the recognition of ADL [2]. This study focused on the architecture defined in [4-6], composed by several steps, such as data acquisition, data cleaning, feature extraction, data fusion and artificial intelligence methods. Based on the literature review, the proposed ADL for the recognition with motion and magnetic sensors are running, walking, going upstairs, going downstairs, and standing.

Based on the data acquired, the feature extraction step should measures several features of the sensors' signal, these are the five greatest distances between the maximum peaks, the Average of the maximum peaks, the Standard Deviation of the maximum peaks, the Variance of the maximum peaks, the Median of the maximum peaks, the Standard Deviation of the raw signal, the Average of the raw signal, the Maximum value of the raw signal, the Minimum value of the raw signal, the Variance of the of the raw signal, and the Median of the raw signal have been extracted from the accelerometer, magnetometer and/or gyroscope sensors available on the off-the-shelf mobile device and the fusion of the data should be a function of the number of sensors available. Thus, the method implemented in the framework for the recognition of the ADL will be adapted with the limitations of the mobile device, including the number of sensors available and the low memory and power processing capabilities.

For the development of the framework for the recognition of the ADL, three types of neural networks were created in order to identify the best framework and neural network for the development of each step of the framework for the recognition of ADL, such as MLP with Backpropagation using the Neuroph framework [14], Feedforward Neural Network with Backpropagation using the Encog framework [15], and Deep Neural Networks (Deep Learning) using DeepLearning4j framework [16], verifying that the Deep Neural Networks achieves better results than others.

Due to the limitations of mobile devices and regarding the results obtained with the method for the recognition of ADL with the accelerometer previously performed, presented in [13], which was verified that the best results were obtained with Deep Neural Networks with $L_2$ regularization and normalized data with an accuracy of 85.89%.

Related to the development of a method for the recognition of ADL, this study proves that the best accuracy are always achieved with Deep Neural Networks with $L_2$ regularization and normalized data, and the data fusion increases the accuracy of the method, reporting an accuracy of 86.49% with the fusion of the data acquired from two sensors (*i.e.,* accelerometer and magnetometer), and 89.51% with the fusion of the data acquired from three sensors (*i.e.,* accelerometer, magnetometer and gyroscope).

On the other hand, MLP with Backpropagation and Feedforward Neural Network with Backpropagation achieves low accuracies, because the networks are overfitting, and this problem may be solved with several hypothesis, these are the stopping of the training when the network error increases for several iterations, the application of dropout regularization, the application of $L_2$ regularization, the application of the batch normalization, or the reduction of the number of features in the neural network.

As future work, the methods for the recognition of ADL presented in this study should be implemented during the development of the framework for the identification of ADL, adapting the method to the number of sensors available on the mobile device, but the method that should be implemented is Deep Neural Networks. The data related to this research is available in a free repository [44].

## Acknowledgements

This work was supported by FCT project **UID/EEA/50008/2013** (*Este trabalho foi suportado pelo projecto FCT UID/EEA/50008/2013*).

The authors would also like to acknowledge the contribution of the COST Action IC1303 – AAPELE – Architectures, Algorithms and Protocols for Enhanced Living Environments.